\documentclass[aps,prl,twocolumn,floatfix,nofootinbib,showpacs,superscriptaddress,tightenlines]{revtex4-1}

\usepackage{graphicx}
\usepackage{float}
\usepackage{epstopdf}
\usepackage{epsfig}
\usepackage{amsmath}
\usepackage{color}
\usepackage{amssymb}   
\usepackage{amsfonts}  
\usepackage{amsbsy}    
\usepackage{multirow} 
\usepackage{pstricks}
\usepackage{slashed}
\usepackage{dcolumn}
\usepackage{xparse}
\usepackage{bm}
\usepackage{hyperref}
\usepackage{soul}
\usepackage{bbm}

\newcommand{\al}{\alpha}
\newcommand{\be}{\beta}

\newcommand{\beq}{\begin{equation}}
\newcommand{\eeq}{\end{equation}}
\newcommand{\ba}{\begin{array}}
\newcommand{\ea}{\end{array}}
\newcommand{\bea}{\begin{align}}
\newcommand{\eea}{\end{align}}
\newcommand{\bi}{\begin{itemize}}
\newcommand{\ei}{\end{itemize}}
\newcommand{\ben}{\begin{enumerate}}
\newcommand{\een}{\end{enumerate}}
\newcommand{\bc}{\begin{center}}
\newcommand{\ec}{\end{center}}
\newcommand{\bl}{\begin{flushleft}}
\newcommand{\el}{\end{flushleft}}
\newcommand{\br}{\begin{flushright}}
\newcommand{\er}{\end{flushright}}

\newcommand{\nn}{\nonumber \\}
\newcommand\Eqn[1]{Eq.~(\ref{#1})}  
\newcommand\Fig[1]{Fig.~\ref{#1}} 

\newcommand{\ie}{{i.e.}}

\newcommand\comment[1]{ \hbox{[{\it Comment suppressed here.}\/]} }
\newcommand\hide[1]{}
\newcommand{\skipover}[1]{}

\begin{document}
\title{A fresh look at the generalized parton distributions of light pseudoscalar mesons}
\author{Zanbin Xing} 
\affiliation{School of Physics, Nankai University, Tianjin 300071, China}
\author{Minghui Ding} 
\affiliation{Helmholtz-Zentrum Dresden-Rossendorf, Bautzner Landstra{\ss}e 400, 01328 Dresden, Germany}
\author{Kh\'epani Raya}
\affiliation{Department of Integrated Sciences and Center for Advanced Studies in Physics, Mathematics and Computation, University of Huelva, E-21071 Huelva, Spain.}
\author{Lei Chang}
\affiliation{School of Physics, Nankai University, Tianjin 300071, China}
\date{\today}
\begin{abstract}
We present a symmetry-preserving scheme to derive the pion and kaon generalized parton distributions (GPDs) in Euclidean space. The key to maintaining crucial symmetries under this approach is the  treatment of the scattering amplitude, such that it contains both the traditional leading-order contributions and the scalar/vector pole contribution automatically, the latter being necessary to ensure the soft-pion theorem. The GPD is extracted analytically via the uniqueness and definition of the Mellin moments and we find that it naturally matches the double distribution; consequently, the polynomiality condition and sum rules are satisfied. The present scheme thus paves the way for the extraction of the GPD in Euclidean space using the Dyson-Schwinger equation framework or similar continuum approaches.
\end{abstract}
\maketitle

{\it Introduction}---
The question of how partons inside hadrons are distributed in momentum and position space has surrounded physicists for decades, and successive attempts have been made to answer this question through both experimental and theoretical methods~\cite{Accardi:2012qut,Anderle:2021wcy,Arrington:2021biu}. A quantity that encodes answers to this  fundamental question is the generalized parton distribution (GPD)~\cite{Ji:1996ek,Radyushkin:1997ki,Muller:1994ses,Goeke:2001tz,Diehl:2003ny,Belitsky:2005qn}, which is non-perturbative and contains information on both the longitudinal-momentum and the transverse spatial distributions. In addition, the GPD is deeply connected to hadron properties~\cite{Mezrag:2022pqk}, for example, lower order moments of the GPD can be linked to hadron form factors and the energy-momentum tensor~\cite{Ji:1996ek}, and so charge and mass distributions, as well as pressure and shear forces inside hadrons~\cite{Polyakov:2002yz}.

Being a non-perturbative quantity, an investigation of the GPD requires a sensible non-perturbative approach. A traditional way to studying the GPD is to derive all physical quantities required in the light-front coordinate system~\cite{Broniowski:2007si,Courtoy:2010qn,Chakrabarti:2020kdc}. However, the non-perturbative quantum chromodynamics (QCD) is commonly formulated in Euclidean space. It is indeed very challenging to study the GPD directly in the light-front coordinate system. Naturally, systematic ways of connecting Euclidean with light-front quantities has been explored. Such is the case of quasi and pseudo distributions~\cite{Shastry:2022obb,Joo:2020spy,Cichy:2018mum}, and the use of the overlap representation of the light-front wave function~\cite{Diehl:2000xz}. The latter is rather promising, since all ingredients can be obtained in a covariant formulation and subsequently projected onto the light-front~\cite{Raya:2021zrz,Albino:2022gzs,Mezrag:2016hnp}. However this limits the domain in which the GPD can be computed and sophisticated methods for extrapolation must be employed~\cite{Chavez:2021llq,Chouika:2017rzs}. 

Alternatively, other methods that make it possible to compute the GPD directly in Euclidean space have been also investigated, such as implementations in lattice QCD~\cite{Ji:2013dva,Ji:2020ect} and in continuum field theory methods~\cite{Courtoy:2010qn,mezrag:tel-01180175}. The discussion here concentrates on the implementation in the continuum field theory approach, \ie, using Dyson-Schwinger equations (DSEs)~\cite{Roberts:2021nhw,Huber:2018ned}. The main idea is to calculate the Mellin moments and then identify the GPD via the uniqueness property. Early explorations concerning the pion parton distribution function (PDF), which is understood as the forward limit of the GPD, already showed the necessity of incorporating contributions beyond the typical impulse approximation diagram~\cite{Chang:2014lva}. For the GPD, the need became more evident, since only in this way problems related to the positivity and polynomiality properties could be avoided~\cite{Mezrag:2014jka,Mezrag:2014tva,Mezrag:2016hnp}.
These ideas have been recently revisited, and a novel perspective is provided in Ref.~\cite{Xing:2022mvk}.
Therein it has been shown that one can directly solve for the dressed meson-meson scattering amplitude, and then use it to calculate meson gravitational form factors. Capitalizing on these recent findings, we observe that it is possible to perform a symmetry-preserving calculation of the GPD using DSEs in Euclidean space. 

In general, the discussion here is universal for all mesons. However, we have special interests in light pseudoscalars, particularly pion and kaon. Contemporary research has shown that there are various connections between the properties of pion and kaon and the emergent hadronic masses (EHM)~\cite{Ding:2022ows}; connections that have been firmly established empirically. Therefore, in this article, we will adopt the symmetry-preserving scheme proposed in~\cite{Xing:2022mvk}, 
\ie, consider the full meson-meson scattering amplitude and use its results, to calculate the Mellin moments of the pion and kaon GPDs; subsequently, the GPDs are then identified from the definition of its Mellin moments.

\begin{figure*}[t]
\includegraphics[width=17.2cm]{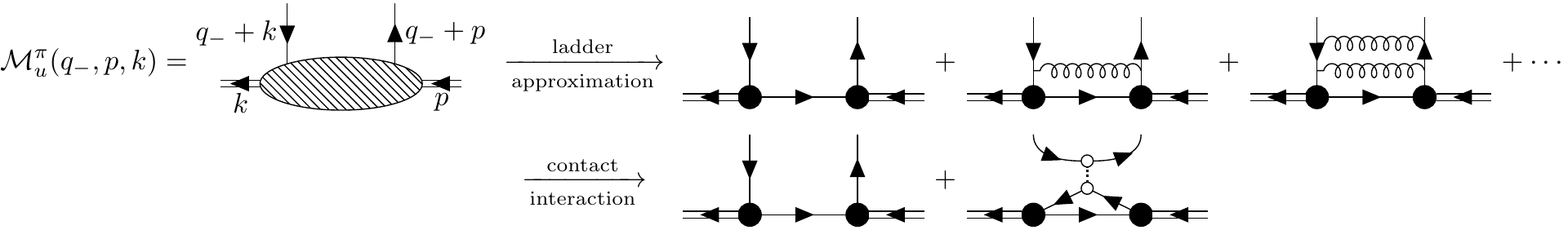}
\caption{Graphical representation of the scattering amplitude $\mathcal{M}^{\pi}_{u}(q_{-},p,k)$. The first row denotes the scattering amplitude under the ladder truncation. The second row denotes the scattering amplitude calculated in a symmetry preserving manner under contact interaction, and the contributions of the two diagrams are $\mathcal{M}^{\pi}_{u,1}(q_{-},p,k)$ and $\mathcal{M}^{\pi}_{u,2}(q_{-},p,k)$ respectively, \ie, $\mathcal{M}^{\pi}_{u}=\mathcal{M}^{\pi}_{u,1}+\mathcal{M}^{\pi}_{u,2}$.  Solid, double-solid, dotted lines represent quark, pion, and the effective scalar/vector meson propagator $\Delta^{s,v}$, respectively. Filled and empty circles represent pion BSA and bare quark-scalar/vector vertex, respectively.}
\label{fig::feyngff}
\end{figure*}

{\it Scattering amplitude: truncation and model}---
We consider as an example the up-quark leading-twist vector GPD in $\pi^{+}$, \ie, $H^{\pi}_{u}(x,\xi,Q^2)$, (whose isospin decomposition can be obtained accordingly), which can be expressed in the form
\begin{equation}\label{eqn::gdp}
 2H^{\pi}_{u}(x,\xi,Q^2)=N_c \text{tr}\int_{q} i\gamma_{n}\mathcal{M}^{\pi}_{u}(q_{-},p,k)\,,
\end{equation}
where $\mathcal{M}^{\pi}_{u}(q_{-},p,k)$ is the off-forward scattering amplitude;
in the Euclidean metric, $p=P-Q/2$ and $k=P+Q/2$ are the momentum of the incoming and outgoing pion, respectively; consequently, $Q$ is the momentum transfer and $P$ is the average momentum; $q_{-}=q-P$; $N_c=3$ is the color degree of freedom and $\text{tr}$ indicates a trace over spinor indices; $\gamma_{n}=\delta_{n}^{x}(q)\gamma\cdot n$ is a generalization of the bare quark-photon vertex $\gamma_\mu$, $\delta_{n}^{x}(q)=\delta(n\cdot q-xn\cdot P)$, and $n$ is a light-like vector, \ie, $n^2=0$; the `skewness', defining the longitudinal momentum transfer, is defined as $\xi=-\frac{n\cdot Q}{2n\cdot P}$ and we focus on the $0<\xi<1$ domain. The range $x\in [\xi,1]$ is the Dokshitzer-Gribov-Lipatov-Altarelli-Parisi (DGLAP) region, while $x\in[-\xi,\xi]$ is the Efremov-Radyushkin-Brodsky-Lepage (ERBL) region.

To obtain the scattering amplitude $\mathcal{M}^{\pi}_{u}(q_{-},p,k)$, a non-perturbative approach is required, such as lattice QCD simulation~\cite{Ji:2013dva,Alexandrou:2020zbe}, and the DSEs framework based on continuum field theory~\cite{Chang:2009zb}. In the framework of DSEs, the leading-order truncation is the ladder approximation, which is sufficient to guarantee charge conservation and the emergence of pions as Goldstone bosons of dynamical chiral symmetry breaking~\cite{Bender:1996bb}. $\mathcal{M}^{\pi}_{u}(q_{-},p,k)$ under the ladder approximation is graphically represented in the first row of Fig.~\ref{fig::feyngff}, which can be interpreted as containing an infinite number of gluon exchange contributions. It can be immediately read from Fig.~\ref{fig::feyngff} that there are two options for calculating the GPD, either dressing $\gamma_{n}$~\cite{Bednar:2018mtf,Mezrag:2014jka} or dressing the scattering amplitude $\mathcal{M}^{\pi}_{u}(q_{-},p,k)$ directly~\cite{Xing:2022mvk}. 

Furthermore, the gluon interaction form in Fig.~\ref{fig::feyngff} is required. Although one can calculate the scattering amplitude directly using realistic gluon models, this is numerically cumbersome. In this article, as the first work to calculate the GPD consistently in the framework of DSEs, we tend to make an attempt first using the simple model~\cite{Frederico:1992np,Gutierrez-Guerrero:2010waf} to demonstrate the process, highlighting that the techniques developed here can be implemented with realistic interaction situations without difficulty.  In the simple model, \ie,  contact interaction, the exchanged gluons between the quarks are approximated as zero-range interactions, $\mathcal{G}_{\mu\nu}(k-q)=\delta_{\mu\nu}/m_{G}^{2}$,
with $m_G$ a gluon mass scale. This framework demands a regularization scheme of the 4-momentum integrals~\cite{Gutierrez-Guerrero:2010waf}, for example those concerning the quark propagator DSE and meson Bethe-Salpeter equation (BSE). We adopt a symmetry preserving regularization approach from Ref.~\cite{Xing:2022jtt} and employ the model parameters listed therein. Under these circumpstances, the momentum-independent nature of the gluon interaction, produces compact expressions for  the quark propagator and the Bethe-Salpeter amplitude (BSA) of the pseudoscalar meson:
\begin{subequations}
\begin{eqnarray}
 S_{f}(k)&=&(-i\gamma\cdot k + M_{f})/D(k)\,,\\
    \Gamma_{\text{PS}}(P)&=&i\gamma_{5}E_{\text{PS}}(P)+\frac{\gamma_{5}\gamma\cdot P}{M_{fg}}F_{\text{PS}}(P)\,,
\end{eqnarray}
\end{subequations}
where $D(k)=k^{2}+M_{f}^{2}$, $M_{fg}=\frac{2M_f M_g}{M_f+M_g}$, $f$ and $g$ denote quark flavors, $M_{f,g}$ is the constituent quark mass;  $E_{\text{PS}}$ and $F_{\text{PS}}$ are two scalar functions that depend only on the total momentum of the pseudoscalar meson. 

The scattering amplitude $\mathcal{M}^{\pi}_{u}(q_{-},p,k)$ can be systematically calculated, the result of which is illustrated in the second row of Fig.~\ref{fig::feyngff}, where the effective scalar/vector meson propagator (the dotted line) can be expressed as
\begin{equation}
    \Delta^{s,v}(Q^{2})=\frac{1}{3m_{G}^{2}/4+f^{s,v}(Q^{2})}\,,
\end{equation}
with $f^{s,v}$ the one-loop scalar/vector vacuum polarization. As with the quark propagator and the pseudoscalar meson BSA, the effective scalar/vector meson propagator can be solved numerically by the consistent rainbow-ladder truncation of the corresponding equation. Up to this point, the scattering amplitude $\mathcal{M}^{\pi}_{u}(q_{-},p,k)$ under the contact interaction is obtained, and calculating the GPD seems straightforward.

{\it Mellin moments and double distributions}---
In Euclidean space, one convention is to start with the Mellin moments of the GPD, which can be written as
\begin{equation}\label{eqn::gdpmmt}
 2\langle x^m \rangle^{\pi}_{u}(\xi,Q^2)
=N_c\text{tr} \int_{q}\frac{(n\cdot q)^m}{(n\cdot P)^{m+1}} i\gamma\cdot n \mathcal{M}_u(q_{-},p,k)\,,
\end{equation}
from the definition $\langle x^m \rangle^{\pi}_{u}(\xi,Q^2)=\int x^{m}H^{\pi}_{u}(x,\xi,Q^2)dx$. For convenience, we label the two components of $\langle x^m \rangle^{\pi}_{u}$ corresponding to the two diagrams in the second row of Fig.~\ref{fig::feyngff} as $\langle x^m \rangle^{\pi}_{u,1}$ and $\langle x^m \rangle^{\pi}_{u,2}$, and similarly the two components of $H^{\pi}_{u}$ are $H^{\pi}_{u,1}$ and $H^{\pi}_{u,2}$. If Mellin moments are calculated, they can then be used to reconstruct the GPD numerically~\cite{Chang:2013pq,Ding:2019qlr} or analytically~\cite{Mezrag:2014tva,Mezrag:2014jka,Mezrag:2016hnp,Xu:2018eii,Raya:2021zrz,Albino:2022gzs}.

We first observe that the denominators of $\langle x^m \rangle^{\pi}_{u,1}$ and $\langle x^m \rangle^{\pi}_{u,2}$ 
can be parameterized respectively as
\begin{subequations}
\begin{eqnarray}
\frac{1}{D_{-Q/2}D_{Q/2}D_{P}}&=&\int_{\Omega_u} \frac{du_1du_2}{[\left(q+\al Q-\be P)^2+\omega_{3}\right]^3}\,,\\
\frac{1}{D_{-Q/2}D_{Q/2}}&=&\int_{\Omega_u} \frac{du_1du_2\delta(\beta)}{\left[(q+\al Q)^2+\omega_{2}\right]^2}\,,
\end{eqnarray}
\end{subequations}
where $D_{X}=D(q-X)$, $\al=u_1-u_2$, $\be=1-u_1-u_2$ and $\Omega_u=\{(u_1,u_2)|0<u_1<1,0<u_2<1-u_1\}$ is the domain of integration in the Feynman parameter space. $\omega_3=\omega(\beta,\alpha)$, $\omega_2=\omega(0,\alpha)$ and
$\omega(\beta,\alpha)=M^2-\bar{\beta} \beta m_\pi^2+\frac{1}{4}Q^2 \left(\bar{\beta}^2-\alpha^2\right)$.
Here we use the isospin symmetry $M_{u/d}=M$, and $\bar{\beta}=1-\beta$.

We then observe that the numerators of $\langle x^m \rangle^{\pi}_{u,1}$ and $\langle x^m \rangle^{\pi}_{u,2}$, when shifting the loop moment $q\to q-\al Q+\be P$, can both be expressed generally as
\begin{equation}
\frac{1}{n\cdot P}\left(\frac{n\cdot q}{n \cdot P}+\beta+\xi\alpha \right)^m \sum_{a,b,c}(n\cdot q)^a (P\cdot q)^b (Q\cdot q)^c f_{abc}(q^2)\,,
\end{equation}
where the coefficient $f_{abc}(q^2)$ is an even function of the momentum $q$, and the powers are restricted by the trace so that $a\in\{0,1\}$ and $b+c\in\{0,1,2\}$. Considering $n^2=0$, the numerators can be simplified by the fact that the integral
\begin{equation}
\binom{m}{j}(\al\xi+\be)^{m-j}\int_{q}(n\cdot q)^{j+a} (P\cdot q)^b (Q\cdot q)^c f_{abc}(q^2)\,,
\end{equation}
survives if and only if 
\begin{equation}
0\leq j+a\leq b+c \,,\quad 
j+a+b+c\in \text{even number}\,.
\end{equation}

By performing the regularization procedure~\cite{Xing:2022jtt} and changing the integration variable $(u_1,u_2)$ to $(\beta,\alpha)$, the integration domain changed accordingly from $\Omega_u$ to $\Omega=\{(\alpha,\beta)|0<\beta<1,-\bar{\beta}<\alpha<\bar{\beta}\}$, we obtain the Mellin moments $\langle x^m \rangle^{\pi}_{u,1}$ and $\langle x^m \rangle^{\pi}_{u,2}$ respectively as
\begin{subequations}
\begin{eqnarray}
\langle x^m \rangle^{\pi}_{u,1}&=&\int_{\Omega}d\beta d\alpha \left[ (\beta+\xi\alpha)^{m}(h_{a0}+\xi h_{a1})\right.\nn
&&\left.+m (\beta+\xi\alpha)^{m-1}(h_{b0}+\xi h_{b1}+\xi^{2} h_{b2})\right]\,,\label{eqn::moment}\\
\langle x^m \rangle^{\pi}_{u,2}&=&\int_{\Omega}d\beta d\alpha \, (\beta+\xi\alpha)^{m}(h_{c0}+\xi h_{c1})\delta(\be)\,,
\end{eqnarray}
\end{subequations}
where the kernel $h_i,i\in\{a0,a1,b0,b1,b2,c0,c1\}$ depends on $(\beta,\alpha,Q^{2})$.
We would like to emphasize that the coefficient function of $(\beta+\xi\alpha)^{m}$ can always be expressed in the linear form of $\xi$, while the coefficient function of $m(\beta+\xi \alpha)^{m-1}$ can always be expressed in the quadratic form of $\xi$, following $h_{a0,b0,b2,c0/a1,b1,c1}$ is the even/odd function of $\alpha$. Furthermore, we would prefer to point out that $h_{c0/c1}$ contains the vector/scalar pole contribution respectively.

The second line of \Eqn{eqn::moment} needs special attention. Noting the fact (we denote $(\beta+\xi\alpha)$ as $(\cdot)$) that $m(\cdot)^{m-1}$ can be written as $\frac{\partial (\cdot)^{m}}{\partial \beta}$ or $\frac{\partial (\cdot)^{m}}{\xi\partial \alpha}$, one can turn $m(\cdot)^{m-1}$ into the form associated with $(\cdot)^{m}$ by using integration by parts. Since the kernel contains two variables, $\alpha$ and $\beta$, the arbitrariness of the differentiation is unavoidable. We choose a particular format  for illustration, namely
\begin{eqnarray}\label{eqn::differentiation}
    & &m(\cdot)^{m-1}(h_{b0}+\xi h_{b1}+\xi^{2} h_{b2})\nn
    &=&\frac{\partial \left[(\cdot)^{m}(h_{b0}+\eta\xi h_{b1})\right]}{\partial\beta}-(\cdot)^{m}\frac{\partial (h_{b0}+\eta\xi h_{b1})}{\partial\beta}\nonumber\\
    &+&\frac{\partial \left[(\cdot)^{m}(\bar{\eta}h_{b1}+\xi h_{b2})\right]}{\partial\alpha}-(\cdot)^{m}\frac{\partial (\bar{\eta}h_{b1}+\xi h_{b2})}{\partial\alpha}\,,
\end{eqnarray}
with the arbitrary parameter $\eta$, and $\eta+\bar{\eta}=1$. 
Substituting \Eqn{eqn::differentiation} into the second line of \Eqn{eqn::moment}, we finally obtain the Mellin moments 
\begin{equation}\label{eqn:DDmmt}
  \langle x^m \rangle^{\pi}_{u}=\int_{\Omega}d\beta d\alpha (\beta+\xi\alpha)^{m}\left[\mathcal{F}(\beta,\alpha,Q^2)+\xi\mathcal{G}(\beta,\alpha,Q^2)\right]\,,
\end{equation}
where
\begin{subequations}
\begin{eqnarray}
    \mathcal{F}(\beta,\alpha,Q^2)&=&h_{a0}-\frac{\partial h_{b0}}{\partial\beta}-\bar{\eta}\frac{\partial h_{b1}}{\partial\alpha}+h_{c0}\delta(\beta)\nn
    &&+h_{b0}\left[\delta(\bar{\beta}-|\alpha|)-\delta(\beta)\right]\nonumber\\
    &&+\bar{\eta}h_{b1}\left[\delta(\bar{\beta}-\alpha)-\delta(\bar{\beta}+\alpha)\right]\,,\\
\mathcal{G}(\beta,\alpha,Q^2)&=&h_{a1}-\eta\frac{\partial h_{b1}}{\partial\beta}-\frac{\partial h_{b2}}{\partial\alpha}+h_{c1}\delta(\beta)\nn
    &&+\eta h_{b1}\left[\delta(\bar{\beta}-|\alpha|)-\delta(\beta)\right]\nonumber\\
    &&+h_{b2}\left[\delta(\bar{\beta}-\alpha)-\delta(\bar{\beta}+\alpha)\right]\,.
\end{eqnarray}
\end{subequations}
This is nothing but the Mellin moments of the double distribution~\cite{Teryaev:2001qm}, and the appearance of $\eta$ exhibits the ambiguity of the double distribution. Comparing \Eqn{eqn:DDmmt} with \Eqn{eqn::gdpmmt}, the GPD is obtained analytically from the double distribution integrated along a line of equation $x-\beta-\xi\alpha=0$, \ie,
\begin{eqnarray}
    H^{\pi}_{u}(x,\xi,Q^2)&=&\int_{\Omega}d\beta d\alpha\,\delta(x-\beta-\xi\alpha) \nonumber\\
    &&\times\left[\mathcal{F}(\beta,\alpha,Q^2)+\xi\mathcal{G}(\beta,\alpha,Q^2)\right]\,.
\end{eqnarray}
 As a complement to the scheme of the $\alpha$ representation of propagators~\cite{Radyushkin:1997ki,Broniowski:2007si}, we have provided a double distribution calculation scheme based on the Feynman technique. 

{\it Discussions}--- In this section, we summarize some properties of the GPD that we have obtained. 

\paragraph{Polynomiality condition:}
Based on the symmetry properties of $h_i$, one would obtain that the double distribution functions 
$\mathcal{F}(\beta,\alpha,Q^{2})=\mathcal{F}(\beta,-\alpha,Q^{2})$
and $\mathcal{G}(\beta,\alpha,Q^{2})=-\mathcal{G}(\beta,-\alpha,Q^{2})$ 
on the domain $\Omega$. Consequently, the Mellin moments fulfill the polynomiality condition and are even functions of $\xi$, with the power of $\xi$ in the polynomial at most $m+1$.

\paragraph{Sum rules:}
Particularly, the zeroth Mellin moment of the GPD corresponds to the electromagnetic form factor
\begin{equation}
 \langle x^0 \rangle^{\text{PS}}_{f}(\xi,Q^2)=F^{\text{em,PS}}_{f}(Q^2)\,,
\end{equation}
and the first Mellin moment corresponds to the gravitational form factors
\begin{equation}
 \langle x^1 \rangle^{\text{PS}}_{f}(\xi,Q^2)=A^{\text{PS}}_{f}(Q^2)+\xi^2D^{\text{PS}}_{f}(Q^2)\,.
\end{equation}
Following the regularization procedure in Ref.~\cite{Xing:2022jtt}, we have analytically checked the equivalence between the form factors calculated in Ref.~\cite{Xing:2022mvk} and the results extracted here using Mellin moments. Thus, the sum rules hold.

\begin{figure}[t]
\includegraphics[width=8.6cm]{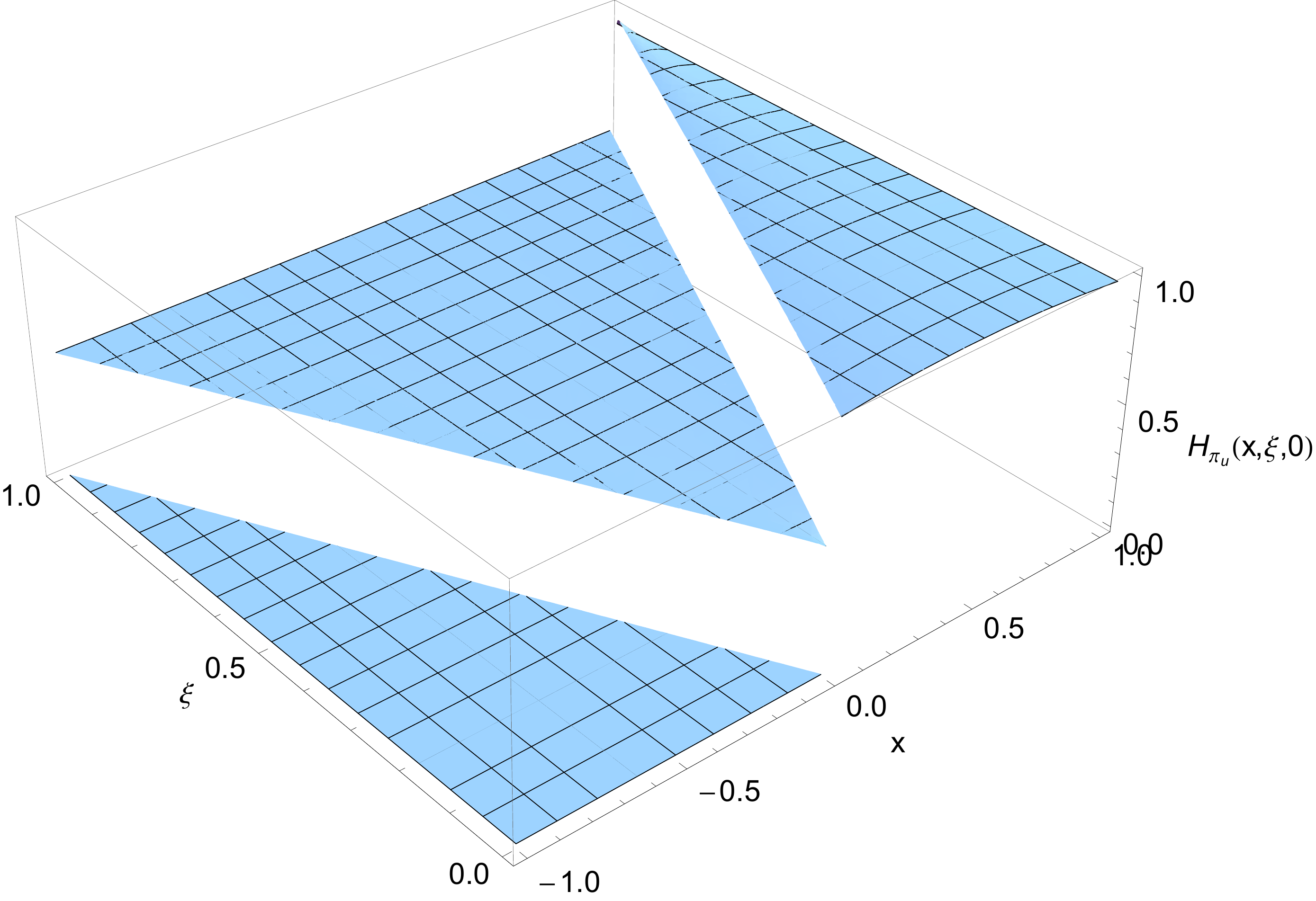}
\includegraphics[width=8.6cm]{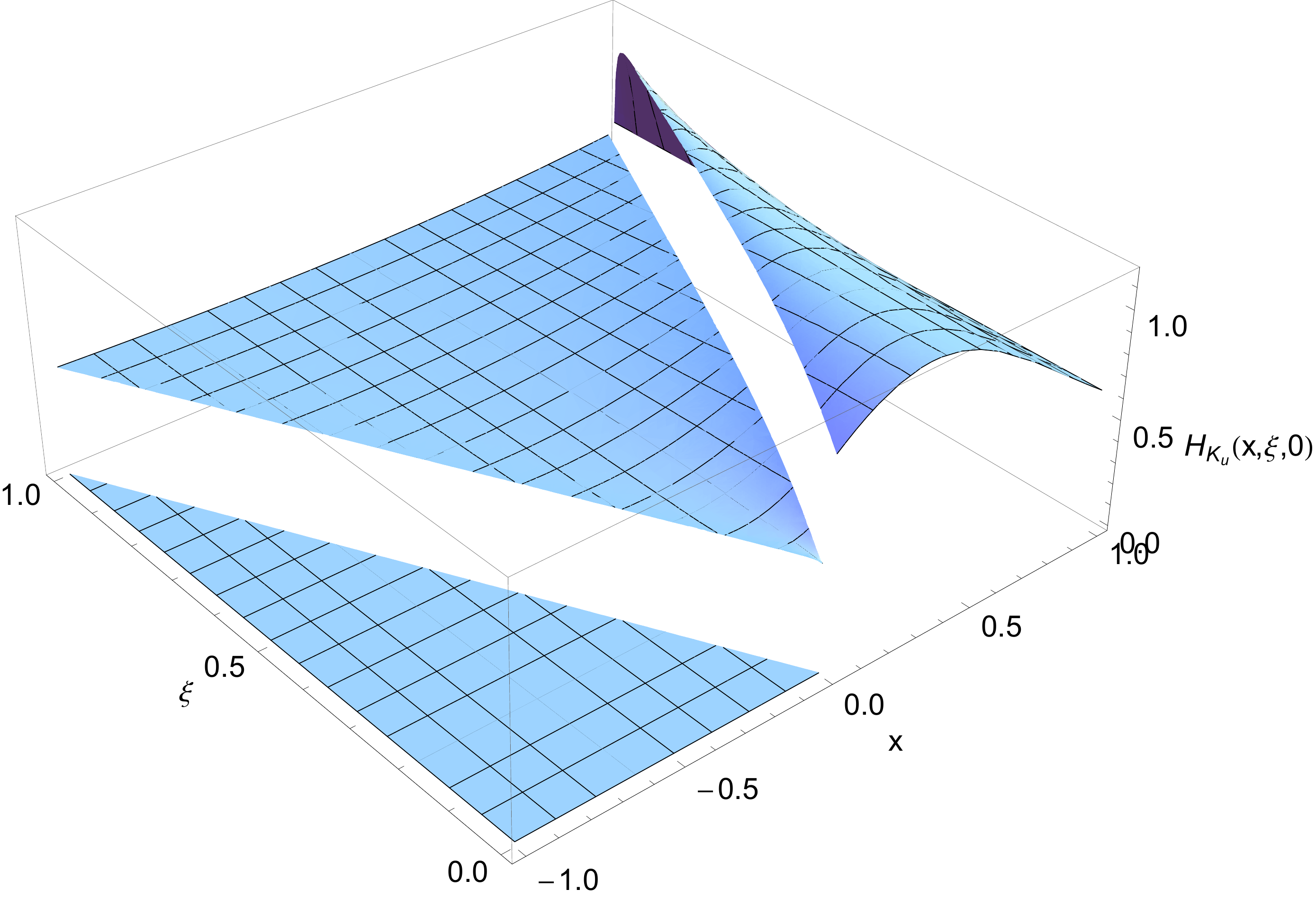}
\caption{Up quark GDPs in pion (upper panel) and kaon (lower panel) with zero momentum transfer $Q^2=0$.}
\label{fig::GPDs}
\end{figure}

\paragraph{Soft-pion theorem:} 
In the chiral limit $m_{\pi}=0$ and at $Q^{2}=0$, the two components of the GPD
corresponding to the two diagrams in the second row of Fig.~\ref{fig::feyngff}
can be written analytically as
\begin{subequations}
\begin{eqnarray}
    H^{\text{c.l.}}_{u,1}(x,\xi,0)&=&\frac{(\xi+x)E_{\text{PS}}-2\xi F_{\text{PS}}}{2\xi(E_{\text{PS}}-2 F_{\text{PS}})}\theta (\xi-x,x+\xi )\nonumber\\
   &&+\theta (1-x,x-\xi )\,,\\
     H^{\text{c.l.}}_{u,2}(x,\xi,0)&=&\frac{-x E_{\text{PS}}}{2\xi(E_{\text{PS}}-2F_{\text{PS}})}\theta (\xi-x,x+\xi )\,,
\end{eqnarray}
\end{subequations}
and added up to
\begin{equation}\label{eqn::softpion}
     H^{\text{c.l.}}_u(x,\xi,0)=\frac{1}{2} \theta (\xi -x,x+\xi )+\theta (1-x,x-\xi )\,.
\end{equation}
From Eq.~\eqref{eqn::softpion} and the fact of constant behavior of parton distribution amplitude, we note that the soft-pion limit at $\xi=1$ is well satisfied. Additionally, it is worth emphasizing that the scalar pole contribution is important for achieving such limit, which cannot be true if only $H^{\text{c.l.}}_{u,1}$ is considered~\cite{Zhang:2020ecj}. 
%


The up-quark GPD in pion and kaon beyond the chiral limit have been depicted in \Fig{fig::GPDs}. We see that $H_{u}^{\pi}(x,0,0)$ is symmetric around $x=1/2$, while $H_{u}^{K}(x,0,0)$ is skewed, peaking at $x<1/2$. The GPD in the $-\xi<x<\xi$ domain, \ie, the ERBL region, is a direct result of this approach, and is positive. \Fig{fig::GPDs} also shows that both GPDs vanish at $x<-\xi$. Additionally, it is noting that  GPDs are discontinuous at $x=\pm\xi$ and non-vanishing at $x=1$, typical results of the contact interaction used here. In the case of realistic interactions, the continuity of GPD is to be expected.



{\it Summary}---
In this work, we show the procedure and numerical results for the calculation of light pseudoscalar meson GPDs in Euclidean space within a consistent scheme.
%
%
The symmetry-preserving treatment of the scattering amplitude is crucial in this computational process, and this treatment allows the scattering amplitude to include not only the contribution of the conventional triangle diagram but also that of an additional diagram containing the scalar/vector meson poles. In doing so, it is seen that the latter is a necessary part that cannot be neglected, otherwise soft-pion theorem is violated. In order to extract the GPD directly in the Euclidean space, we evaluate the corresponding Mellin moments and, by using Feynman parameters and other algebraic approaches, we systematically identify the GPD and obtain the corresponding double distribution. Specifically, there are several novel findings in this computational process: $(i)$ The symmetry-preserving regularization scheme allows us to include all terms after tracing without the need to reduce the common factors in the numerator and denominator in the GPD Mellin moment formula, as was commonly performed in Ref.~\cite{Broniowski:2007si,mezrag:tel-01180175}. $(ii)$ In this process, a form of $m(\cdot)^{m-1}$ arises, and different handles of this term can reflect the ambiguity of the double distribution. $(iii)$ Since the GPD we obtained matches the double distribution, the symmetry restrictions required by QCD, such as the polynomiality condition and sum rules are satisfied. 
Finally, it is worth noting that our approach is universal for all sorts of mesons, and although we use the contact interaction model to illustrate all the computational processes of light pseudoscalar mesons, the present approach opens a window for studying meson GPDs with sophisticated interactions.

{\it Acknowledgments}---
Work supported by National Natural Science Foundation of China (grant no. 12135007). MD is grateful for support by Helmholtz-Zentrum Dresden-Rossendorf High Potential Programme.

\bibliography{main}
\end{document}